\documentclass[a4paper,10pt,pre,twocolumn,showpacs,aps,floats,superscriptaddress]{revtex4}

\usepackage{epsfig}

\begin{document}

\title{Dynamics of jamming transitions in complex networks}

\author{Pablo Echenique}

\affiliation{Departamento de F\'{\i}sica Te\'orica, Universidad de
Zaragoza, Zaragoza 50009, Spain.}

\affiliation{Instituto de Biocomputaci\'on y F\'{\i}sica de Sistemas
Complejos, Universidad de Zaragoza, Zaragoza 50009, Spain}

\author{Jes\'us G\'omez-Garde\~nes}

\affiliation{Departamento de F\'{\i}sica de la Materia Condensada,
Universidad de Zaragoza, Zaragoza 50009, Spain}

\affiliation{Instituto de Biocomputaci\'on y F\'{\i}sica de Sistemas
Complejos, Universidad de Zaragoza, Zaragoza 50009, Spain}

\author{Yamir Moreno}

\affiliation{Instituto de Biocomputaci\'on y F\'{\i}sica de Sistemas
Complejos, Universidad de Zaragoza, Zaragoza 50009, Spain}

\affiliation{Departamento de F\'{\i}sica Te\'orica, Universidad de
Zaragoza, Zaragoza 50009, Spain.}

\date{\today}

\widetext

\begin{abstract} 

We numerically investigate jamming transitions in complex
heterogeneous networks. Inspired by Internet routing protocols, we
study a general model that incorporates local traffic information
through a tunable parameter. The results show that whether the
transition from a low-traffic regime to a congested phase is of first
or second order type is determined by the protocol at work. The
microscopic dynamics reveals that these two radically different
behaviors are due to the way in which traffic jams propagate through
the network. Our results are discussed in the context of Internet
dynamics and other transport processes that take place on complex
networks and provide insights for the design of routing policies based
on traffic awareness in communication systems.

\end{abstract}


\maketitle

Everyday, transportation networks$-$ airports, roads, the Internet,
etc$-$ carry on a huge amount of load in the form of passengers,
vehicles or information packets delivered by millions of users when
searching the World-Wide-Web (WWW), sending and receiving e-mails, or
looking at stock market quotes. The globalization and the Information
Era have in turn led to a continuous growth of most communication
networks driven by the increase in traffic; capital examples being the
Internet and main airports. The efficient performance of these systems
is mainly determined by the ability of the system to avoid congestion
and reduce transit times. Congestion on networked systems often comes
up suddenly and provokes the breakdown of the system's
functionality. Hence, it is not surprising that transport processes
and information exchange have been widely studied over the last years
because of potential applications in fields as diverse as sociology
\cite{ws98}, urban planning (vehicular traffic)
\cite{nagel93,nagel95}, informatics \cite{ohira98,tak00,t2,sv04,us04}
and biology \cite{tlalka03}.

A great body of work in the subject has been carried out for regular
and random graphs. Real networks are however complex. The patterns of
interconnections describing the interactions of the system's elements
have been unraveled just a few years ago. Surprisingly, in sharp
contrast with the common sense, systems as diverse as the Internet,
the WWW, biological and social networks \cite{strogatz,book1,book2}
share a number of topological features. This fact makes the modeling
of complex networks an attractive field as one may define and study
general models and then translate the conclusions and implications to
the language of a particular field. For instance, the resilience to
random failures and attacks and the spreading of virus and rumors
\cite{alexvb,havlin01,newman00,moreno02,av03,n02b} are relevant
problems in sociology, biology and technological systems.

The intense research on complex networks during the last years have
provided deep insights into the dependency of the properties of
dynamical processes on the topological properties of the underlying
networks in which these processes take place. Of particular interest
are the so-called scale-free (SF) networks \cite{bar99}, since they
have been shown to radically change well-established results for
random graphs and regular lattices. These networks are ubiquitous in
Nature. They are made up of a number of nodes (or elements) and the
probability that a given node has $k$ connections to other nodes
follows a power law $P_k\sim k^{-\gamma}$. This is the case of the
Internet which shows a scale-free (SF) connectivity distribution with
an exponent that has been estimated to be around $\gamma=2.2$
\cite{alexvb}.

In this paper, we study the impact of traffic routing protocols on the
performance of communication systems when traffic awareness is
incorporated. More specifically, we are interested in exploring how
the average network performance depends on the ability of the routing
protocol to divert traffic across paths other than the shortest
ones. To this end, we numerically explore a model in which a tunable
parameter accounts for the degree of traffic awareness incorporated in
packets delivery. We find that the onset of the congested phase is
reminiscent of a first or a second order phase transition depending on
whether or not the routing combines a shortest path delivery strategy
with traffic aggregation at a local scale. We also report on the
differences in local dynamics and discuss our results in the context
of cost-performance trade-offs associated with different routing
strategies. 

In order to be more precise, we shall discuss in what follows our
model and results within the field of technological networks such as
the Internet. However, as we will see, the model is aimed at
simulating a general transport process on top of a complex scale-free
network. Hence, our results could also be useful in other
fields. Moreover, as it was recently shown \cite{us04}, different
routing strategies are sensitive to local details of the networks
under study$-$ for instance, to the degree-degree correlations and the
clustering coefficient$-$, so that it is advisable to use real
networks. Hence, we have used the Internet Autonomous System map
at the Oregon route server dated May 25, 2001 \cite{as}, which is a SF
network with $\gamma=2.2$ and $N=11174$ nodes.

The model is defined as follows. Starting from an unloaded network, we
impose a constant input of newly created packets. At each time step,
$p$ information packets are created. The source of each packet as well
as its destination are chosen at random among all the nodes of the
network. Besides, each node sends only one packet at each time step
and, as a consequence, one node, $i$, can have a queue of $c_{i}$
packets to be delivered. The path-delivery strategy is sketched
below. Let us suppose that a node $l$ is holding a packet whose
destination is $j$. Then, the effective distance between a neighbor
$i$ ($i=1,\ldots,k_l$) of $l$ and $j$ is defined as
\begin{equation}
\delta_i=hd_i+(1-h)c_i,
\label{eq0}
\end{equation}
where $d_i$ is the minimum number of hops one has to pass by in order
to reach $j$, i.e., the shortest path between $i$ and
$j$. Additionally, $h$ is a tunable parameter that accounts for the
degree of traffic awareness incorporated in the delivery algorithm. It
is worth noting that when $h=1$, we recover a shortest-path delivery
protocol similar to actual Internet routing mechanisms$-$ in the
Internet, routers deliver data packets by converging to a best
estimate of the path connecting each destination address$-$ or to the
typical behavior in urban traffic. The packet is finally diverted
following the path going through the node that minimizes
$\delta_i$. Henceforth, we will distinguish between the case $h=1$,
referred to as {\em standard} protocol, and $h\neq 1$, called {\em
traffic-aware} scheme. Other strategies are also possible \cite{us04},
but it has been recently shown that the two described above give the
best network's performance when packets are only introduced at the
beginning of the process.

\begin{figure}[t]
\begin{center}
\epsfig{file=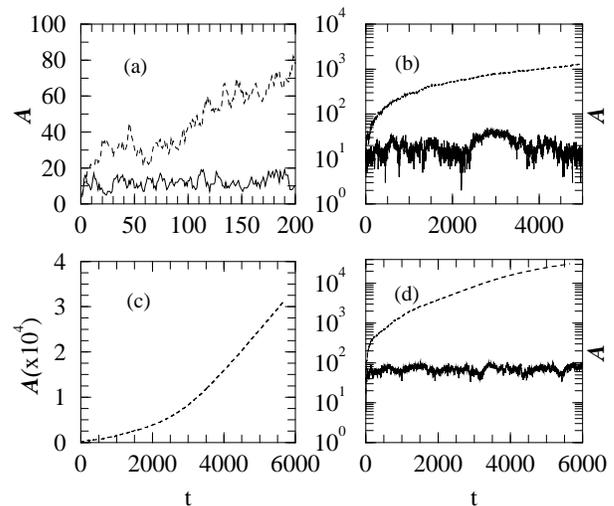,width=2.7in,angle=-90,clip=1}
\end{center}
\caption{Total number of active packets as a function of time
steps. Figures $(a)$ and $(b)$ correspond to the standard protocol,
while $(c)$ and $(d)$ have been obtained for the traffic-aware routing
with $h=0.85$. In each figure, the continuous line stand for
subcritical values of $p$ ($(a)$ and $(b)$ $p=3.0$, $(d)$ $p=8.0$) and
the dotted line corresponds to $p>p_c$ ($(a)$ and $(b)$ $p=4.0$, $(c)$
and $(d)$ $p=13.0$). See the text for further details on the
definitions.}
\label{figure1}
\end{figure}

The procedure for $h\neq 1$ combines knowledge of the structural
properties of the network and its current dynamical state at a local
scale. It allows to divert packets through larger but less congested
paths, consequently a trade-off associated to packets' transit times
is naturally and dynamically incorporated. As an appropriate measure
of the efficiency of the process, we monitor the aggregation of
packets in the network, given by the number of packets that have not
reached their destinations at each time step $t$, $A(t)$. Figure\
\ref{figure1} shows the results obtained for different values of $p$
and $h$. As it can be seen, when the external driving is applied at
low rates (i.e., small $p$), both protocols allow for a stationary
state. In this state, the system is able to balance the in-flow of
packets with the flow of packets that reach their destinations. The
stationary state, where no macroscopic signs of congestion is
observed, corresponds to a free flow phase. The situation changes when
the rate at which new packets are introduced increases. As we will see
below, there is a {\em critical} value $p_c$ beyond which a congested
phase shows up. Let us now note that for the standard protocol (Fig.\
\ref{figure1}a, dotted line), when $p>p_c$, $A(t)$ grows linearly in
time $\forall t$. On the contrary, for the traffic-aware algorithm, we
observe that $A(t)$ grows slowly at short times and then becomes
steeper as time goes on with a constant slope (Fig.\ \ref{figure1}c).

In order to characterize the phase transition from a free phase to a
congested one, we introduce the order parameter
\begin{equation}
\rho=\lim_{t\to\infty}\frac{A(t+\tau)-A(t)}{\tau
p},
\label{eq1}
\end{equation}
where $\tau$ is the observation time. The limit in Eq.\ (\ref{eq1}) is
introduced only to ensure that the system is not in a temporary
regime. Equation\ (\ref{eq1}) hence measures the ratio between the
outflow and the inflow of packets during a time window $\tau$. $\rho$
equals $1$ when the congestion is maximal (no packet reaches its
destination) and $0$ when an equilibrium is established, i.e., in the
stationary state.

\begin{figure}[t]
\begin{center}
\epsfig{file=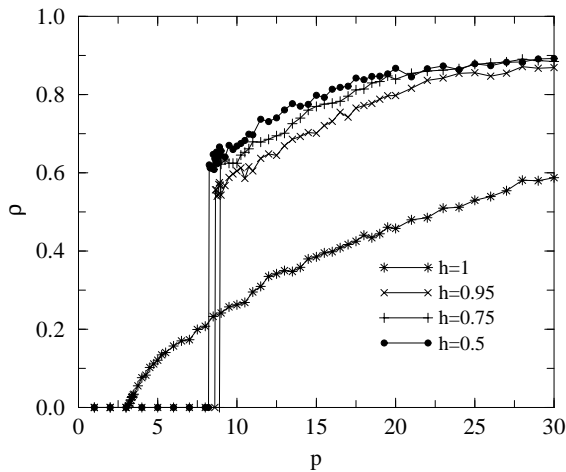,width=2.5in,angle=-90,clip=1}
\end{center}
\caption{Jamming transitions as a function of $p$. The order parameter
  $\rho$ is given by Eq.\ (\ref{eq1}). Note that $h=1$ corresponds to
  the standard strategy in which traffic awareness is absent. As soon
  as traffic conditions are taken into account, the jamming transition
  is reminiscent of a first-order phase transition and the critical
  point shifts rightward.}
\label{figure2}
\end{figure}

Figure\ \ref{figure2} depicts the system's phase diagram. The dynamics
of the system is characterized in both protocols by a critical point
beyond which a macroscopic congestion arises. However, there are two
radically different behaviors for the onset of traffic jams. In the
standard protocol ($h=1$), the critical point is small, $p_c=3$ and
the jamming transition is reminiscent of a second order phase
transition. On the contrary, when $h\neq 1$, the critical point $p_c
\approx 9$ is distinctly larger than for $h=1$, but the appearance of
a congested phase turns out to be consistent with a first order phase
transition, with a sharp jump of $\rho$ at the transition
point. Moreover, the order of the transition for the latter protocol is
independent of $h$ provided that $h\neq 1$.

\begin{figure}[t]
\begin{center}
\epsfig{file=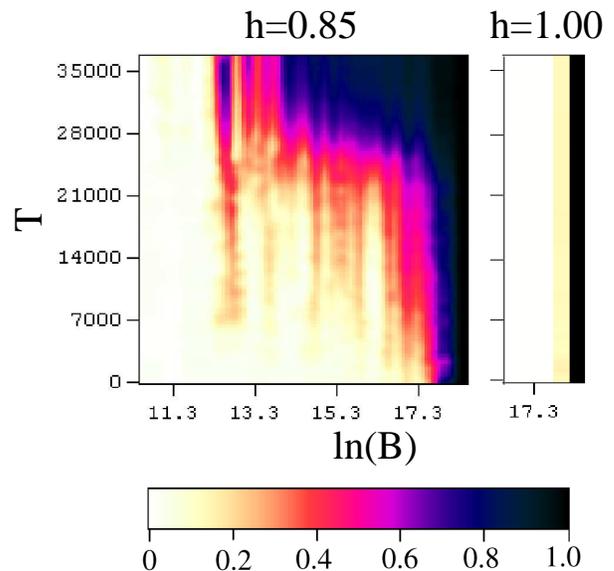,width=3.1in,angle=0,clip=1}
\end{center}
\caption{(Color online) Congestion levels as a function of time and
  nodes' betweenness. At each time step, the color-coded scale is
  normalized by the number of packets $c_i$ in the queue of the node
  with the largest congestion. Two radically distinct behaviors are
  obtained for the standard ($h=1$, $p=4>p_c=3$, right panel) and the
  traffic-aware ($h=0.85$, $p=13>p_c=9$, left panel) protocols.}
\label{figure3}
\end{figure}

The two different types of transitions depending on whether or not
traffic-awareness is incorporated in the protocol at work, poses an
interesting issue. Which of the two protocols will be best suited to
handle traffic? It depends on the system. While for the standard
protocol we get a smaller critical point, the jammed phase does not
appear suddenly. Hence, if we would like to have a system in which
traffic jams appear and grow smoothly, the standard algorithm is the
best choice. On the contrary, we could implement a sort of
traffic-aware protocol if we are interested in delaying the appearance
of congestion, however at the cost of a sudden jump to a highly jammed
phase due to the lack of previous warnings.

In order to provide more insights into the nature of the phase
transitions, we now focus on the microscopic details of the system's
dynamics. We have inspected how the nodes get congested. As both
protocols incorporate a shortest path delivery strategy, a suitable
description can be obtained by monitoring the number of active packets
at each node as a function of the betweenness of the nodes$-$ which,
on the other hand, scales with $k$ \cite{alexvb}. The betweenness or
load of a node $i$ gives the total number of shortest paths among all
pairs of nodes in the network that pass through $i$
\cite{alexvb,new01,goh}. It is a measure of the centrality of a node
in the network and becomes a relevant quantity in traffic flow
modeling. Figure\ \ref{figure3} clearly illustrates the distribution
of congested vertices for the two protocols analyzed. The shortest
paths connecting the sources and the destinations of any active packet
always tend to visit first the more connected nodes and then go down
to the less connected ones. This is a consequence of the hierarchy of
the network and is called up-down strategy \cite{alexvb}. For $h=1$,
the protocol only works on a shortest path delivery basis. The hubs
become congested early in the process causing the packets to get
trapped in a few nodes as shown in Fig.\ \ref{figure3}. When traffic
conditions are taken into account by the routing mechanisms, the same
up-down strategy applies up to the hubs. Then, instead of getting
trapped in them, the packets circumvent highly jammed nodes and the
load is distributed to nodes other than the hubs, provoking the
aggregation of traffic in neighborhoods of overcrowded nodes. In the
long time limit, the congestion is spread through the network (see,
Fig.\ \ref{figure3}) shifting the critical point to larger values of
$p$.

\begin{figure}[t]
\begin{center}
\epsfig{file=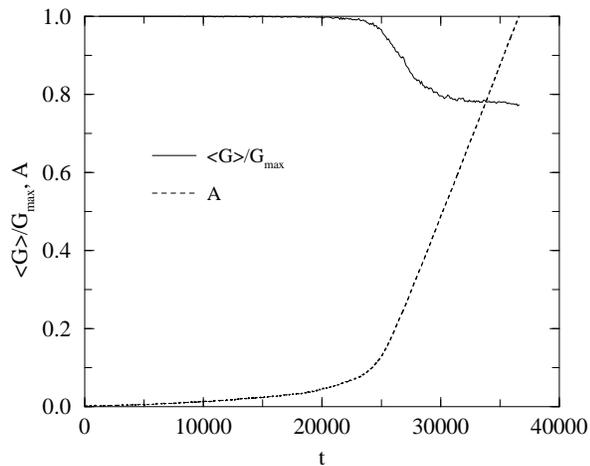,width=2.5in,angle=-90,clip=1}
\end{center}
\caption{Time dependence of the total number of packets in the system
  and average size of the clusters formed by non impermeable
  nodes. Note that $A(t)$ becomes steeper just when the inflection of
  $\langle G \rangle/G_{max}(t)$ changes. $h=0.85$ and $p=13$. See the
  text for further details. }
\label{figure4}
\end{figure}

It is possible to get deeper into what is going on in the system for $h\neq 1$.
Let us suppose again that a node $l$ is holding a packet to be sent to $j$
through one of its neighbors $i$ ($i=1,\ldots,k_l$). Among all the neighbors of $l$,
there is one node with the lowest load $c_{min}$. Now, assume the extreme
situations in which by going through $i$ the packet is one hop closer to its
destination, but taking the path for which $c_i=c_{min}$, it is one hop farther
from $j$. Thus it follows that whenever the relation $c_i-c_{min}>2h/(1-h)$ is
verified, the packet will never be sent through $i$. This node $i$ is
impenetrable for $l$. If a node is impenetrable for all its neighbors, we call
it just {\em impermeable}, since it does not participate in traffic delivery.
As congestion spreads throughout the network, the number of impermeable nodes
increases and changes dynamically.  Therefore a dynamical backbone made up of
all nodes that are able to transmit the packets comes up.  The picture is
similar to the percolation of a fluid through a porous media.  Here, packets
can flow only through non impermeable nodes as a fluid can only flow through
the pore channels.

The existence of impermeable nodes provokes the appearance of both
small network components in the form of impenetrable regions, and
clusters of {\em allowed} paths. Figure\ \ref{figure4} depicts the
time dependence of the average cluster size (normalized by the largest
cluster size) of allowed regions. Starting from $t=0$, as time goes
on, the total number of packets in the network increases and there is
only one cluster of the size of the network. When signs of congestion
first appear, $\langle G \rangle/G_{max} (t)$ decreases departing from
unity. At longer times, traffic jams reach more nodes (see, Fig.\
\ref{figure3}, for $t>21000$) causing the congestion to be more
distributed in the network. Finally, the flow of packets in the
network reaches a regime in which $A(t)$ increases linearly in time
and $\rho$ saturates to its stationary value. In this state, marked by
an inflection point in the $\langle G \rangle/G_{max} (t)$ curve
beyond which the average cluster size of allowed regions stabilizes,
the system seems to have self-organized the distribution of jammed
nodes. This self-organization phenomenon nicely explains why one can
not go from one protocol to the other by making $h=1$, as it can seem
from Eq.\ (\ref{eq0}). The discontinuity at $h=1$ is therefore due to
the lack of alternative paths in the standard protocol. Even for $h$
very close to $1$, the system will self-organize itself into a state
in which congested nodes are distributed and not limited to the very
hubs of the network. The only dependence with $h$ is manifested in the
time needed for self-organization, that becomes very large and
eventually diverge when $h\to 1$.

In conclusion, we have characterized jamming transitions in complex
heterogeneous networks. The results show that when traffic awareness
is incorporated into the routing protocol, new cooperative effects
arise. Additionally, the jamming transitions are well described by two
radically different phase transitions. Finally, our results
demonstrate that whether or not a given protocol is best suited for
traffic handling depends on a delicate trade-off between the system's
performance and traffic capabilities (how large $p_c$ is) and how
congestion arises (smoothly or suddenly). The model thus provides
useful insights for the design of new routing policies and may be a
guide for more complex models where, for instance, routers can tune
$h$ dynamically depending on the traffic conditions at a local scale.

\begin{acknowledgments}
P.\ E.\ and J.\ G-G\ acknowledge financial support of the MECyD
through FPU grants. Y.\ M.\ is supported by MEyC through the Ram\'on y
Cajal Program. This work has been partially supported by the Spanish
DGICYT projects BFM2002-01798 and BFM2002-00113.
\end{acknowledgments}

\end{document}